\documentclass[onecolumn,newstyle]{rmaa}
\nonstopmode

\usepackage[latin1]{inputenc}
\usepackage{latexsym}

\ReceivedDate{} 
\AcceptedDate{} 

\title{An Efficient Monte Carlo Algorithm for a \\Restricted
Class of Scattering Problems\\ in Radiation Transfer}

\author{Alan~M.~Watson and William~J.~Henney
\affil{Instituto de Astronomía, Unidad Morelia, Universidad
Nacional Autónoma de México} }

\shortauthor{Watson \& Henney}
\shorttitle{Monte Carlo Radiation Transfer}

\fulladdresses{\item Alan M.\ Watson and William J.\ Henney:
Instituto de Astronomía, Unidad Morelia, Universidad
Nacional Autónoma de México, Apartado Postal 72-3 (Xangari),
58089 Morelia, Michoacán, México
(a.watson, w.henney@astrosmo.unam.mx).}

\resumen{\ignorespaces Se describe un algoritmo eficiente
tipo Monte Carlo para solucionar una clase restringida de
problemas en la transferencia de radiación. Esta clase
incluye varios problemas de interés en la astrofísica,
incluyendo la dispersión de la luz ultravioleta y visible
por granos. El algoritmo toma en cuenta la dispersión
múltiple. Se describe el algoritmo, se presentan algunas
optimizaciones importantes, y se muestra explícitamente cómo
se puede usar el algoritmo para estimar cantidades como la
intensidad emergente y promedio. Se presenta dos pruebas del
algoritmo, se examina la importancia de las optimizaciones,
y se muestra que el algoritmo se puede aplicar de manera
útil a los problemas ópticamente delgados, los cuales son a
veces considerados limitados a aproximaciones explícitas de
dispersión sencilla más atenuación.}

\abstract{\ignorespaces We describe an efficient Monte Carlo
algorithm for a restricted class of scattering problems in
radiation transfer. This class includes many astrophysically
interesting problems, including the scattering of
ultraviolet and visible light by grains. The algorithm
correctly accounts for multiply-scattered light. We describe
the algorithm, present a number of important optimizations,
and explicity show how the algorithm can be used to estimate
quantities such as the emergent and mean intensity. We
present two test cases, examine the importance of the
optimizations, and show that this algorithm can be usefully
applied to optically-thin problems, a regime sometimes
considered limited to explicit single-scattering plus
attenuation approximations.

\bigskip}

\keywords{Radiative Transfer --- 
Scattering --- Methods:~Numerical}

\renewcommand{\vec}[1]{\mathbf#1}
\renewcommand{\r}{\vec r}
\newcommand{\n}{\vec n}

\renewcommand{\[}{\begin{eqnarray}}
\renewcommand{\]}{\end{eqnarray}}

\begin{document}


\maketitle



\section{Introduction}
\label{section:introduction}

We present a Monte Carlo algorithm for the numerical
solution of a restricted class of scattering problems in
radiation transfer. This class consists of problems in which
the true emissivity (the emissivity of unscattered light)
and the opacity, albedo, and scattering properties are
specified a priori and are zero outside a finite domain, but
have arbitrary distributions within that finite domain. In
particular, no restrictions are placed on the scattering
phase function or on the symmetry of the problem. Examples
of problems that satisfy these requirements include those in
which the opacity is dominated some combination of grain
opacity, electron scattering, resonance-line scattering,
Rayleigh scattering, or Raman scattering. For example, the
transfer of ultraviolet and visible light in the presence of
grains and of Ly$\alpha$ photons in the presence of neutral
hydrogen and grains. For simplicity, in this paper we do not
consider time-dependent problems or polarization, however,
as we discuss in \S\ref{section:extensions}, the algorithm
can be easily extended to cover these cases.

We have developed this algorithm over the last several
years, initially independently (Watson 1994; Henney 1994ab;
Henney 1995; Henney \& Axon 1995; Burrows et al.\ 1996;
Sahai et al.\ 1998; and Henney 1998) and then
collaboratively (Stapelfeldt et al.\ 1999; Watson et al.\
2001). The algorithm currently incorporates several
important features and optimizations, including the use of
forced scatterings and forced interactions, the construction
of estimators from unbiased samples of scattering events,
and a very efficient integrator for optical depth.

Monte Carlo algorithms have been widely used in radiation
transfer (see \S\ref{section:previous-work}). It seems clear
to us that aspects of this algorithm have been independently
discovered and appear to be common knowledge among those
working in this field. The principal contribution of this
paper is to formalize and publish the algorithm, so that it
can be better understood and so that future researchers can
avoid the wasted effort of independent rediscovery.

In \S2 we present the restricted problem along with a
formal (but computationally intractable) solution. In
\S3 we briefly discuss previous approaches to this
problem. In \S4 we present the algorithm and a number
of optimizations and variations. In \S5 we discuss
important details of our implementations. In \S6 we
present two test cases. In \S7 we quantify the efficiency
gains due to two important optimizations. In \S8 we
briefly outline how the algorithm might be extended to
include time dependence and polarization. In \S9 we
summarize our main results.


\section{The Restricted Problem}

\begin{table*}[tp]
\caption{Notation}
\label{table:notation}
\begin{center}
\begin{tabular}{lll}
\hline
\hline

Symbol&
Definition&
Units
\\

\hline

$\r$&
Position.&
cm
\\

$\r_i$&
Position of the $i$-th interaction with matter.&
cm\\

$\n$&
Direction; $|\n| \equiv 1$.&
\\

$\n_i$&
Direction after the $i$-th interaction with matter.&
\\

$\nu$&
Frequency.&
Hz
\\

$\nu_i$&
Frequency after the $i$-th interaction with matter.&
Hz\\

$w$&
Statistical weight.&
\\

$w_i$&
Statistical weight after the $i$-th interaction with matter.&
\\

$\eta(\r, \nu, \n)$&
Total emissivity; $\sum_{i=0}^{\infty}\eta_i(\r, \nu, \n)$.&
$\mathrm{erg\,s^{-1}\,cm^{-3}\,sr^{-1}\,Hz^{-1}}$\\

$\eta_i(\r, \nu, \n)$&
Partial emissivity of light scattered $i$
times.&
$\mathrm{erg\,s^{-1}\,cm^{-3}\,sr^{-1}\,Hz^{-1}}$\\

$I(\r, \nu, \n)$&
Total specific intensity; $\sum_{i=0}^{\infty}I_i(\r, \nu, \n)$.&
$\mathrm{erg\,s^{-1}\,cm^{-2}\,sr^{-1}\,Hz^{-1}}$\\

$I_i(\r, \nu, \n)$& Partial specific intensity of light scattered $i$ times.&
$\mathrm{erg\,s^{-1}\,cm^{-2}\,sr^{-1}\,Hz^{-1}}$\\

$\mathcal{L}(\r, \nu, \n)$&
Total radiative intensity, defined in \S\ref{section:L};
$\sum_{i=0}^{\infty}\mathcal{L}_i(\r, \nu, \n)$.&
$\mathrm{erg\,s^{-1}\,sr^{-1}\,Hz^{-1}}$\\

$\mathcal{L}_i(\r, \nu, \n)$& Partial radiative intensity of light scattered $i$ times.&
$\mathrm{erg\,s^{-1}\,sr^{-1}\,Hz^{-1}}$\\

$l$&
Length.&
cm\\

$\tau(\r, \nu, \n; l)$&
Optical depth from $\r$ to $\r + l \n$;
$\int_{0}^l dl'\:
\chi(\r + l' \n, \nu, \n)
$.\\

$\tau_\infty(\r, \nu, \n)$&
Optical depth from $\r$ to infinity;
$\lim_{l\rightarrow \infty} \tau(\r, \nu, \n; l)$.\\ 

$\chi(\r, \nu, \n)$&
Linear extinction coefficient; $\kappa +
\sigma$&
$\mathrm{cm^{-1}}$\\

$\sigma(\r, \nu, \n)$&
Linear scattering coefficient.&
$\mathrm{cm^{-1}}$\\

$\kappa(\r, \nu, \n)$&
Linear absorption coefficient.&
$\mathrm{cm^{-1}}$\\

$a(\r, \nu, \n)$&
Single-scattering albedo; $\sigma/\chi$.\\

$\Sigma(\r; \nu', \n'; \nu, \n)$& 
Scattering outcome function, defined by equation
\ref{equation:sigma-definition}.&
$\mathrm{sr^{-1}\,Hz^{-1}}$\\

$\Phi(\r; \nu; \n', \n)$&
Scattering phase function, normalized according to equation
\ref{equation:Phi-normalization}.&
$\mathrm{sr^{-1}}$\\

\hline
\hline

\end{tabular}
\end{center}
\end{table*}

The restricted problem requires that $\eta_0(\r, \nu, \n)$,
$\chi(\r, \nu, \n)$, $a(\r, \nu, \n)$, and $\Sigma(\r; \nu',
\n'; \nu, \n)$ be specified a priori.
Table~\ref{table:notation} defines our notation, which
closely follows that of Mihalas (1978), except for the
scattering phase function: we use $\Phi$ instead of $g$ for
the phase function, to avoid confusion with the asymmetry
parameter of the Henyey-Greenstein phase function, and we
normalize the phase function according to equation
\ref{equation:Phi-normalization}, so the phase function has
units of $\mathrm{sr^{-1}}$ and an isotropic phase function
has a value of $1/4\pi~ \mathrm{sr^{-1}}$ everywhere. Note
that the emissivities and specific intensities are per unit
solid angle and the extinction, scattering, and absorption
coefficients are per unit length. We denote by $x_i$ the
quantity $x$ when only light scattered $i$ times is
considered. For example, $\eta_0$ is the emissivity of
unscattered light, that is, the true emissivity, and
$\eta_1$ is the emissivity of singly-scattered light. For
convenience, we also introduce the single-scattering albedo
$a$ and the scattering outcome function $\Sigma$. The albedo
$a$ has its conventional definition as the probability that
an interaction with matter results in a scattering rather
than an absorption. The scattering outcome function $\Sigma$
is defined by
\[
\label{equation:sigma-definition}
\eta(\r, \nu, \n) = \eta_0(\r, \nu, \n)
+
\int_0^\infty \!\!\!\! d\nu' \!\!
\int_{4\pi} \!\!\!\! d\Omega' \!\!
\;\;
\sigma(\r, \nu', \n')
\Sigma(\r; \nu', \n'; \nu, \n)
I(\r, \nu', \n')
\]
That is, $\Sigma(\r; \nu', \n'; \nu, \n)$ is the
contribution to the emissivity $\eta(\r, \nu, \n)$ when
intensity $I(\r, \nu', \n')$ is scattered. We explicitly
separate $\sigma$ and $\Sigma$ to distinguish the
probability distribution of scattering events from the
probability distribution of the outcomes of those events
given that they have occurred.

Scattering does not necessarily conserve energy, but it
conserves photons. Considering the specific intensity as
being carried by photons of energy $h\nu$, we can identify
$\Sigma(\r; \nu', \n'; \nu, \n) (\nu'/\nu)$ as the joint
probability distribution that upon scattering a photon with
frequency $\nu'$ and direction $\n'$ becomes a photon with
frequency $\nu$ and direction $\n$. Thus, integrating over
all initial or final states, we have
\[
\int_0^\infty \!\!\!\! d\nu' \!\!
\int_{4\pi} \!\!\!\! d\Omega' \!\!
\;\;
\Sigma(\r; \nu', \n'; \nu, \n)
(\nu'/\nu)
=
\int_0^\infty \!\!\!\! d\nu \!\!
\int_{4\pi} \!\!\!\! d\Omega \!\!
\;\;
\Sigma(\r; \nu', \n'; \nu, \n)
(\nu'/\nu)
\equiv
1,
\]
for any $\nu$ and $\n$ or $\n'$ and $\n'$. As an example of
$\Sigma$, consider coherent scattering ($\nu'= \nu$) with a
phase function $\Phi(\r; \nu; \n', \n)$ normalized as
\[
\label{equation:Phi-normalization}
\int_{4\pi} \!\!\!\! d\Omega' \!\!
\;\;
\Phi(\r; \nu; \n', \n) 
=
\int_{4\pi} \!\!\!\! d\Omega \!\!
\;\;
\Phi(\r; \nu; \n', \n) 
\equiv
1;
\]
we then have
\[
\Sigma(\r; \nu', \n'; \nu, \n)
(\nu'/\nu)
= 
\delta(\nu' - \nu) \Phi(\r; \nu; \n', \n).
\]

The standard formal solution of
the equation of radiation transfer gives the intensity
$I_i(\r, \nu, \n)$ of photons that have undergone $i$
scatterings as
\[
\label{equation:transfer}
I_i(\r, \nu, \n) = 
\int_{l=-\infty}^{0} \!\!\!\!\!\!\!\!\!\! dl \!\!
\;\;
\eta_i(\r + l\n, \nu, \n)
e^{-\tau(\r, \nu, \n; l)},
\]
where
\[
\tau(\r, \nu, \n; l) = \int_{0}^l\!\! dl'\:
\chi(\r + l' \n, \nu, \n).
\]
To evaluate the first integral, we need the emissivities
$\eta_i(\r, \nu, \n)$. The true emissivity $\eta_0(\r, \nu,
\n)$ of unscattered light is specified a priori. The
emissivity $\eta_i(\r, \nu, \n)$ of light that has undergone
$i$ scatterings (where $i > 0$) is given by
\[
\label{equation:scattering}
\eta_i(\r, \nu, \n) =
\int_0^\infty \!\!\!\! d\nu' \!\!
\int_{4\pi} \!\!\!\! d\Omega' \!\!
\;\;
\sigma(\r, \nu', \n') 
\Sigma(\r; \nu', \n'; \nu, \n) 
I_{i-1}(\r, \nu', \n').
\]
In general, equations \ref{equation:transfer} and
\ref{equation:scattering} form an infinite sequence of
coupled integral equations. Approximate solutions can be
obtained by directly evaluating a few terms and ignoring the
rest. The single-scattering-plus-attenuation approximation
consists of directly evaluating $I_1$, the intensity of
single-scattered light, and ignoring multiply-scattered
light ($I_2$, $I_3$, etc.); this
is often adequate for optically-thin problems. However, in
optically-thick problems, multiply-scattered light can be
important, yet direct calculation of the intensity $I_i$ for
a single set of values of $\vec r$, $\nu$, and $\vec n$
requires the evaluation of a $3i+1$ dimensional integral, and
rapidly becomes intractable; the Monte Carlo algorithm
addresses this intractability.


\section{Previous Work}
\label{section:previous-work}

The restricted scattering problem described in the preceding
section has been tackled many times using semi-analytic and
Monte Carlo methods. 

Semi-analytic methods include the discrete-ordinate method
(Chandrasekhar 1960; Flannery, Roberge, \& Rybicki 1980;
Stamnes et~al.\ 1988) and ``doubling'' and ``adding''
methods (van de Hulst 1963; Hansen \& Travis 1974; de Haan,
Bosma, \& Hovenier 1987). They are well-suited to problems
with plane-parallel or spherical geometries, but are very
difficult to extend to arbitrary geometries. Additionally,
these methods become less efficient when the scattering
phase function is sharply peaked (Escalante 1994), as
appears to be the case for interstellar grains in the
visible and ultraviolet.

Monte Carlo methods, in which photon trajectories are
simulated probabilistically, offer more flexibility. Witt
(1977), Hillier (1991), Whitney (1991), Fischer, Henning, \&
Yorke (1994), Watson (1994), Code \& Whitney (1995), Henney
\& Axon (1995), and Bjorkman (1997) have presented
descriptions of various Monte Carlo methods, and Cashwell \&
Everett (1959) have described general techniques that can
greatly increase the efficiency of Monte Carlo methods.
Monte Carlo methods have been applied to astrophysical
problems many times over the last several decades; the
earliest application we can find is an investigation of the
transfer of resonance-line radiation in plane-parallel slabs
by Avery \& House (1968). Faster computers have allowed
increasingly complicated geometries to be explored in recent
years.





\section{The Algorithm}
\label{section:algorithm}

The algorithm consists of two logically separate parts. Part
I follows the course of many pseudo-photons as they scatter
through the system, producing a sample of ``interaction
events''. These interactions events are characterized by the
quantities $\r_i$, $\nu_i$, and $\n_i$ (the location of the
$i$-th interaction with matter and the frequency and
direction \emph{after} the $i$-th interaction with matter)
and by a statistical weight $w$. We first present a naive
algorithm for Part I that straightforwardly simulates the
underlying physical processes, and then present several
variations and optimizations, which trade computational
efficiency for conceptual complexity. Part II makes Monte
Carlo estimates of derived quantities, such as the emergent
and mean intensities. It is based on the observation that
the weighted values of the interaction events generated in
Part I form unbiased samples of the photon emissivities
$\eta_i/h\nu$ and the combinations $I_i\chi/h\nu$. This
algorithm extends those we have seen published in its use of
forced scatterings, forced interactions, and, in particular,
its approach to estimating the emergent intensity (``forced
escapes'').

\subsection{Normalization}

We begin by considering the normalization of the problem.
The general radiation transfer problem is non-linear, as it
includes coupling of radiation and matter. However, the
restricted problem is linear, and we can impose a convenient
normalization on $\eta_0$ and all derived quantities. Again,
considering emissivity in terms of photons of energy $h\nu$,
we can identify $\eta_0(\r, \nu, \n)/h\nu$ as being
proportional to the joint probability distribution that a
photon is emitted from position $\r$ with frequency $\nu$
and direction $\n$. We can thus impose a normalization such
that
\[
\int_\infty \!\!\!\! dV \!\!
\int_0^\infty \!\!\!\! d\nu \!\!
\int_{4\pi} \!\!\!\! d\Omega \!\!
\;\;
\eta_0(\r, \nu, \n) / h\nu
\equiv 1,
\]
so that $\eta_0(\r, \nu, \n)/h\nu$ is identically the joint
probability distribution of photon emission. All derived
quantities are thus implicitly normalized by the number of
emitted photons.


\subsection{Part I: Generating The Sample of Interaction
Events (Naive Algorithm)}

The sample of interaction events is constructed by
simulating the emission and transfer of an adequately large
number of pseudo-photons; the values of $w$, $\r_i$,
$\nu_i$, and $\n_i$ for each pseudo-photon form the sample.
The algorithm is followed for each pseudo-photon until the
pseudo-photon escapes or is absorbed. We first describe a
naive version which is a direct analog of the transfer of a
real photon (considered as a particle) through the real
system.

\begin{enumerate}

\item[1:] Choose the initial position $\r_0$, frequency
$\nu_0$, and direction $\n_0$ from the joint probability
distribution $\eta_0(\r_0, \nu_0, \n_0) / h\nu_0$.

\item[2:] Initialize the weight: $w \leftarrow 1$.

\item[3:] Initialize the scattering index: $i \leftarrow 0$.

\item[4:] Choose an optical depth $\tau_i$ from an
exponential distribution with a mean of 1.

If $\tau_i$ is greater than the optical depth to infinity
$\tau_\infty(\r_i, \nu_i, \n_i)$, then the pseudo-photon
escapes: stop.

\item[5:] Otherwise, the pseudo-photon interacts with
matter. The interaction occurs after a distance $l_i$ which
is given by the implicit equation
\[
\label{equation:l-i}
\int_{0}^{l_i} dl'\: \chi(\r_i + l' \n_i, \nu, \n_i) 
\equiv
\tau(\r_i, \nu_i, \n_i; l_i) 
= 
\tau_i.
\]

\item[6:] $\r_{i + 1} \leftarrow \r_i + l \n_i$.

\item[7:] Choose a probability $u_i$  from a uniform
distribution between 0 and 1.

If $u_i > a(\r_{i+1}, \nu_i, \n_i)$ then the pseudo-photon
is absorbed: stop.

\item[8:] Otherwise, the pseudo-photon is scattered. Choose a
frequency $\nu_{i+1}$ and direction $\n_{i+1}$ 
from the joint probability distribution $\Sigma(\r_{i+1};
\nu_i, \n_i; \nu_{i+1}, \n_{i+1}) (\nu_i / \nu_{i+1})$.

\item[9:] Increment the scattering index: $i \leftarrow i + 1$.

\item[10:] Continue from step 4.

\end{enumerate}

\subsection{Part I: Generating The Sample of Interaction
Events (Variations and Optimizations)}

We now present two variations and two optimizations. The two
variations can simplify the algorithm, easing its
implementation, at a cost of increasing the variance of the
results. The two optimizations can drastically reduce the
variance in the results at the cost of a small increase in
complexity. (This is demonstrated and discussed in
\S\ref{section:optimizations}). The principal conceptual
change in these variations and optimizations is that the
statistical weight $w$ of the pseudo-photon ceases to be
constant. The basis of the variations is weight balancing.
An event $E$ which should be selected from a probability
distribution $p(E)$ can be selected from a different
probability distribution $p'(E)$ provided $p'(E)$ is
non-zero whenever $p(E)$ is non-zero and the statistical
weight of the event is multiplied by a factor $p(E)/p'(E)$.
This technique is useful as it allows an unwieldy
probability distribution to be replaced by a more
straightforward one. The basis of the optimizations is
weight splitting. An event $E$ with statistical weight $w$
can be replaced by two or more events $E_j$ with statistical
weights $w_j = w p(E_j|E)$ provided that the $E_j$ fully
cover the possible outcomes, that is $\sum p(E_j|E) = 1$.
This technique can be used to reduce the variance in the
sample of events.

\subsubsection{Variation: Choosing $\r_0$, $\nu_0$, and $\n_0$}

In step 1 the initial values of $\r_0$, $\nu_0$, and $\n_0$
are selected from a probability distribution that can be
quite unwieldy. As an alternative, the values can be
selected from different (more easily manageable)
distributions and then the pseudo-photon can be weighted
appropriately. For example, step 1 can be replaced by:

\begin{enumerate}

\item[$1'$:] Choose an initial position $\r_0$ from a uniform
distribution whose domain includes all positions $\r$ for
which $\eta_0(\r, \nu, \n)$ is non-zero for some frequency
$\nu$ and direction $\n$.

Choose an initial frequency $\nu_0$ from a uniform
distribution whose domain includes all frequencies $\nu$ for
which $\eta_0(\r, \nu, \n)$ is non-zero for some position
$\r$ and direction $\n$.

Choose an initial direction $\n_0$ from a uniform
distribution whose domain includes all directions $\n$ for
which $\eta_0(\r, \nu, \n)$ is non-zero for some position
$\r$ and frequency $\nu$.

Initialize the statistical weight: $w \leftarrow \eta_0(\r_0,
\nu_0, \n_0) / h\nu_0$.

\end{enumerate}

Obviously, variations on this are possible if, for example,
some variables can be selected easily from a distribution
but others cannot. All that is required is that the product
of the joint probability density for selection and the
initial statistical weight $w$ be equal to the true joint
probability density for emission $\eta_0(\r_0, \nu_0,
\n_0)/h\nu_0$.


\subsubsection{Variation: Choosing $\nu_{i+1}$ and $\n_{i+1}$}

Again, in step 8 the values of $\nu_{i+1}$ and $\n_{i+1}$
are selected from a probability distribution that can be
quite unwieldy. As an alternative, the values can be
selected from different (more easily manageable)
distributionsq and then the pseudo-photon can be weighted
appropriately. For example, step 8 can be replaced by:

\begin{enumerate}

\item[$8'$:] Choose a frequency $\nu_{i+1}$ from a uniform
distribution whose domain includes all frequencies $\nu$ for
which $\Sigma(\r_{i+1}; \nu_i, \n_i; \nu, \n)$ is non-zero
for some direction $\n$.

Choose a direction $\n_{i+1}$ from a uniform distribution
whose domain includes all directions $\n$ for which
$\Sigma(\r_{i+1}; \nu_i, \n_i; \nu, \n)$ is non-zero for
some frequency $\nu$.

Adjust the statistical weight: $w \leftarrow w
\Sigma(\r_{i+1}; \nu_i, \n_i; \nu_{i+1},
\n_{i+1})(\nu_i/\nu_{i+1})$.

\end{enumerate}

Again, other variations are possible, provided once more
that the product of the joint probability density for
selection and the factor modifying the statistical weight
$w$ is equal to the true joint probability distribution for
scattering $\Sigma(\r_{i+1}; \nu_i, \n_i; \nu_{i+1},
\n_{i+1})(\nu_i/\nu_{i+1})$.


\subsubsection{Optimization: Forced Scatterings}
\label{section:forced-scatterings}

If forced scatterings are required, step 7 is replaced by
the following if $i$ is less than some tuneable parameter:

\begin{enumerate}

\item[$7'$:] Split the pseudo-photon into one pseudo-photon
that scatters and another that is absorbed. Each
pseudo-photon has the same set of values of $\r_j$, $\nu_j$,
and $\n_j$ for $j \le i$ and the same $\r_{i+1}$.

For the pseudo-photon that is absorbed: set the statistical
weight to $w \leftarrow (1-a(\r_{i+1}, \nu_i, \n_i)) w$ and
stop.

For the pseudo-photon that is scattered: set the statistical
weight to $w \leftarrow a(\r_{i+1}, \nu_i, \n_i) w$ and
continue.

\end{enumerate}

The effect of this optimization is that the ramifications of
both scattering and absorption are fully explored; the
pseudo-photon makes appropriately weighted contributions to
both the absorbed and scattered events. This is particularly
important when the albedo is low (see
\S\ref{section:optimizations}).


\subsubsection{Optimization: Forced Interactions}
\label{section:forced-interactions}

If forced interactions are desired, step 4 of the algorithm
is replaced by the following if $i$ is less than a tuneable
parameter:

\begin{enumerate}

\item[$4'$:]

Define the escape probability of the photon along its
current trajectory as $\beta_i \equiv e^{-\tau_\infty(\r_i, \nu_i,
\n_i)}$.

Choose a deviate $\tau_i$ from an exponential distribution
truncated at $\tau_\infty(\r_i, \nu_i, \n_i)$ (that is,
$p(\tau_i) = e^{-\tau_i} / (1 - \beta_i)$ for $0 < \tau_i
\le \tau_\infty(\r_i, \nu_i, \n_i)$ and $p(\tau_i) = 0$
otherwise).

Split the pseudo-photon into one pseudo-photon that escapes
and another that interacts with matter. Each pseudo-photon
has the same set of values of $\r_j$, $\nu_j$, and $\n_j$
for $j \le i$.

For the pseudo-photon that escapes: set the statistical
weight to $w \leftarrow \beta_i w$ and stop.

For the pseudo-photon that interacts: set the statistical
weight to $w \leftarrow (1 - \beta_i) w$ and continue.

\end{enumerate}

The effect of this optimization is that the ramifications of
both escapes and further interactions are fully explored;
the pseudo-photon makes appropriately weighted contributions
to both the escaping and interacting events. This is
particularly important when the system is optically thin
(see \S\ref{section:optimizations}).

%
%
%
%
%
%
%


\subsection{Part II: Estimates of Derived Quantities}

The sample of interaction events can be used to estimate
physically interesting quantities, such as the emergent and
mean intensity. It is clear from the similarity between the
algorithm for Part I and the underlying physical process
that (a) the weighted values of $\r_i$, $\nu_i$, and $\n_i$
are drawn from the joint probability distributions of a
photon being emitted or scattered at position $\r_i$ with
frequency $\nu_i$ into direction $\n_i$ after $i$ scatters
and (b) the weighted values of $\r_{i+1}$, $\nu_i$, and $\n_i$
are drawn from the joint probability distributions of a
photon being absorbed or scattered at position $\r_{i+1}$
after $i$ previous scatters whilst having frequency $\nu_i$
and traveling in direction $\n_i$. These probability
distributions are just $\eta_i(\r, \nu, \n)/h\nu$ and
$I_i(\r, \nu, \n)\chi(\r, \nu, \n)/h\nu$; thus the weighted
values of $\r_i$, $\nu_i$, and $\n_i$ form unbiased samples
of $\eta_i(\r, \nu, \n)/h\nu$ and the weighted values of
$\r_{i+1}$, $\nu_i$, and $\n_i$ form unbiased samples of
$I_i(\r, \nu, \n)\chi(\r, \nu, \n)/h\nu$.

We can use standard Monte Carlo techniques on these unbiased
samples. If we have a function $f(\r,\nu, \n)$ defined over
a volume $V$, a frequency interval $N$, and a solid angle
$\Omega$, then we have
\[
W^{-1}
\!\!\!\!\!\!\!\!\!\!\!\!\!\!\!\!
\sum_{\forall(\r_i, \nu_i, \n_i) \in (V, N, \Omega)} 
\!\!\!\!\!\!\!\!\!\!\!\!\!\!\!\!
w f(\nu_i, \r_i, \n_i) \approx
\int_V \!\!\!\! dV \!\!
\int_N \!\!\!\! d\nu \!\!
\int_\Omega \!\!\!\! d\Omega \!\!
\;\;
[\eta_i(\r, \nu, \n)/h\nu]
f(\r, \nu, \n)
\label{equation:monte-carlo-emissivity}
\]
and
\[
W^{-1}
\!\!\!\!\!\!\!\!\!\!\!\!\!\!\!\!\!\!
\sum_{\forall(\r_{i+1}, \nu_i, \n_i) \in (V, N, \Omega)} 
\!\!\!\!\!\!\!\!\!\!\!\!\!\!\!\!\!\!
w f(\r_{i+1}, \nu_i, \n_i) 
\approx
\int_V \!\!\!\! dV \!\!
\int_N \!\!\!\! d\nu \!\!
\int_\Omega \!\!\!\! d\Omega \!\!
\;\; 
[I_i(\r, \nu, \n)\chi(\r, \nu, \n)/h\nu]
f(\r, \nu, \n),
\label{equation:monte-carlo-extinctivity-i}
\] 
or equivalently
\[
W^{-1}
\!\!\!\!\!\!\!\!\!\!\!\!\!\!\!\!\!\!\!\!\!
\sum
_{\forall(\r_i, \nu_{i-1}, \n_{i-1}) \in (V, N, \Omega)} 
\!\!\!\!\!\!\!\!\!\!\!\!\!\!\!\!\!\!\!\!\!
w f(\r_i, \nu_{i-1}, \n_{i-1}) 
\approx
\int_V \!\!\!\! dV \!\!
\int_N \!\!\!\! d\nu \!\!
\int_\Omega \!\!\!\! d\Omega \!\!
\;\; 
[I_{i-1}(\r, \nu, \n)\chi(\r, \nu, \n)/h\nu]
f(\r, \nu, \n).
\label{equation:monte-carlo-extinctivity-i-1}
\] 
Here, $W \equiv \sum w$ is the total statistical weight of
all pseudo-photons. The approximations are in the sense that
the integrals are the means of the sums and, by the central
limit theorem, are their limiting values as the number of
pseudo-photons tends to infinity. Thus, the sums can be used
to estimate the integrals. As examples, we derive below
expressions for the emergent intensity and the mean
intensity; other quantities such as the heating rate and
radiation pressure can be derived similarly.


\subsubsection{Emergent Specific Intensity}

We define a volume $V$ by the translation of a surface $\vec
S$ in a direction $\n$ from $+\infty$ to $-\infty$. We
define $\bar I_i(\vec S, \nu, \vec n)$ as the average
emergent specific intensity of light scattered $i$ times in
direction $\n$ from volume $V$. It is given by
\[
\bar I_i(\vec S, \nu, \n)
=
(\vec S \cdot \n)^{-1}
\int_V \!\!\!\! dV \!\!
\;\;
\eta_i(\r, \nu, \n)
e^{-\tau_\infty(\r, \nu, \vec n)}.
\]
For the scattered emergent intensity, $\bar I_i(\vec S, \nu,
\n)$ with $i > 0$, we can substitute equation
\ref{equation:scattering} for $\eta_i$ to give
\[
\bar I_i(\vec S, \nu, \n)
=
(\vec S \cdot \n)^{-1}
\int_V \!\!\!\! dV \!\!
\int_0^\infty \!\!\!\! d\nu' \!\!
\int_{4\pi} \!\!\!\! d\Omega' \!\!
\;\;
I_{i-1}(\r, \nu', \n') \sigma(\r, \nu', \n') \Sigma(\r; \nu',
\n'; \nu, \n)
e^{-\tau_\infty(\r, \nu, \n)}
\]
We can now use equation
\ref{equation:monte-carlo-extinctivity-i-1} (with $N$
covering all frequencies and $\Omega$ covering $4\pi$
steradians) to estimate $\bar I_i(\vec S, \nu, \n)$ by
\[
\bar I_i(\vec S, \nu, \n) \approx
(\vec S \cdot \n)^{-1} W^{-1}
\!\!\!\!\!\!\!\!\!\!\!\!\!\!\!
\sum_{\forall (\r_i, \nu_i, \n_i) \in (V, N, \Omega)}
\!\!\!\!\!\!\!\!\!\!\!\!\!\!\!
w 
\,
h\nu_{i-1}
\,
a(\r_i, \nu_{i-1}, \n_{i-1}) \Sigma(\r_i; \nu_{i-1},
\n_{i-1}; \nu, \n) e^{-\tau_\infty(\r_i, \nu, \n)},
\]
where, because $N$ and $\Omega$ are all-inclusive, the
condition on the sum simplifies to $\forall \r_i \in V$.

In some Monte Carlo algorithms, the emergent intensity is
estimated by binning the photons that escape in step 4 of
the algorithm in Part I. Our method of estimating the
emergent intensity is very different, in that it is based on
the set of interaction events generated in steps 5 to 8. For
spherically symmetric problems, the two approaches are
similar in efficiency. However, for other problems our
approach is more efficient by a factor of roughly $4\pi$
divided by the size of the bin in solid angle used in the
other method; thus, our approach can easily be an order of
magnitude more efficient for axisymmetric problems and two
orders of magnitude more efficient for asymmetric problems,
provided the viewing direction is fixed. By analogy with
forced scatterings and forced interactions, we refer to this
approach as ``forced escapes''.

We must use a different approach for the unscattered
emergent intensity $\bar I_0$, as equation \ref{equation:scattering} is
valid only for $i > 0$ and neither
equation \ref{equation:monte-carlo-emissivity} nor
equation \ref{equation:monte-carlo-extinctivity-i} will give $\bar
I_0$ in general. One simple approach is performing
a logically separate Monte Carlo estimation. That is,
selecting values of $\r_0'$ from a uniform distribution that
includes all positions for which $\eta_0(\r, \nu, \n)$ is
non-zero for some $\nu$ and $\n$ and  estimating $\bar
I_0$ by
\[
\bar I_0(\vec S, \nu, \n) \approx
(\vec S \cdot \n)^{-1} 
\left(
\sum_{\r_0' \in V}
\eta_0(\r_0', \nu, \n) e^{-\tau_\infty(\r_0', \nu, \n)}
\over
\sum \eta_0(\r_0', \nu, \n)/h\nu
\right).
\]
Other Monte Carlo schemes are possible and can take
advantage of knowledge of the properties of $\eta_0$. The
two logically separate Monte Carlo estimations can be
combined under certain circumstances.

\subsubsection{Emergent Radiative Intensity}
\label{section:L}

When the system is unresolved, for whatever reason, the
specific intensity can no longer be measured. In this case,
a more useful quantity is the radiative intensity
$\mathcal{L}$, which in more familiar terms is the
luminosity per unit solid angle $\mathcal{L} \equiv
dL/d\Omega$, where $L$ is the total luminosity of the
system. This quantity is defined and used more often in
physics than astronomy, although even in physics is it not
common. The radiative intensity is related to the observed
physical flux $F$ by $\mathcal{L} = Fd^2$, where $d$ is the
distance to the source, and to the specific intensity $I$ by
$\mathcal{L} = \int_S I dS = S \bar I(S),$ where the surface
$S$ includes the whole system. Like the specific intensity,
the radiative intensity has the convenient property of being
independent of distance. We can estimate $\mathcal{L}$ using
the same equations as for the specific intensity, but
letting the volume of integration extend over the whole
system and omitting the division by the area $\vec{S} \cdot
\n$.


\subsubsection{Mean Intensity}

The mean value $\bar J_i(V, N)$ of the mean intensity in a
volume $V$ and frequency range $N$ of light scattered $i$
times is
\[
\bar J_i(V, N)
&=&
{1 \over VN}
\int_V \!\!\!\! dV' \!\!
\int_N \!\!\!\! d\nu' \!\!
\;\;
J_i(\r', \nu')
\\
&=&
{1 \over 4 \pi VN}
\int_V \!\!\!\! dV' \!\!
\int_N \!\!\!\! d\nu' \!\!
\int_{4\pi} \!\!\!\! d\Omega'
\;\;
I_i(\r', \nu', \n')
.
\]
We can use equation \ref{equation:monte-carlo-extinctivity-i} to
estimate $\bar J_i$ by
\[
\bar J_i(V, N) \approx
{1 \over 4 \pi VN}
\sum_{\forall(\nu_i, \r_{i+1}) \in (N, V)}
{
w h\nu_i
\over
\chi(\r_{i+1}, \nu_i, \n_i)
}
.
\]
Note that while $\bar I_i$ is obtained at a single frequency
$\nu$, $\bar J_i$ is obtained in a frequency range $N$. This
method of calculating the mean intensity seems to be widely
used by other Monte Carlo methods, and this particular
formulation does not enjoy any advantage of efficiency.

\subsection{Estimation of Uncertainty}

In addition to calculating the value of a derived quantity
from the whole set of pseudo-photons, we can partition the
set into $M$ equal sub-sets and obtain $M$ independent
values. The uncertainty in the value calculated from the
whole set can be estimated as the standard deviation of the
$M$ values from the $M$ sub-sets divided by $\sqrt M$.


\section{Implementation}

\subsection{Specifying a Problem}

One of the advantages of Monte Carlo methods is their
generality and flexibility. For example, to specify a
problem, our implementations require only five subroutines:
to emit a pseudo-photon, to determine the extinction
coefficient, to determine the albedo, to determine the
location of the sentinel surfaces (see below), and to
scatter a pseudo-photon.

\subsection{Unifying Part I and Part II of the Algorithm}

The algorithm as described above generates a sample of
interaction events in Part I and then derives quantities
from weighted sums of functions of these interaction events
in Part II. Since a simulation can require many millions of
pseudo-photons, a naive implementation would require the
storage of many millions of interaction events. Our
implementations unify Parts I and II, forming the sums as
the interaction events are generated. This complicates the
implementations, but relieves them of the need to store the
entire sample of interaction events.

\subsection{The Integrator for Optical Depth}

There are many ways to solve equation \ref{equation:l-i} for
the distance $l_i$ to the optical depth $\tau_i$, but we
have found it convenient and efficient to treat it as a
differential equation and integrate. That is, we calculate
increments $\Delta l$ and $\Delta \tau$ to $l$ and $\tau$
and iterate until we satisfy the boundary condition $\tau =
\tau_i$, in which case we have solved for $l = l_i$. We use
a fourth-order Runga-Kutta integrator, which in this case
reduces to Simpson's rule. Our integrator uses step doubling
to estimate the fractional and absolute errors and adapts
the step size appropriately. If a step fails, it is tried
again using a mid-point integrator; this requires no further
integrand evaluations and so is cheap; furthermore, it
allows the integrator to integrate up to and away from
discontinuities. We have found that this integrator is
faster than higher-order Runga-Kutta or Gaussian integrators
for the extinction distributions we have encountered. We
believe that this is because we typically require fractional
precisions of only $10^{-5}$ or less. The situation is
reversed if much higher precisions are imposed.

We integrate in a piece-wise manner, breaking the
integration at ``sentinel surfaces'' which correspond to
discontinuities and sharp features such as the equatorial
plane of a thin disk. Thus, we are typically integrating
until we satisfy one of two boundary conditions: until
either $\tau = \tau_i$ or $l = L$, where $L$ is the distance
to the next sentinel surface. If the first condition is
satisfied, we stop the integration; if the second is
satisfied, we calculate the distance to the next sentinel
surface and continue. This approach allows the integrator to
handle discontinuities gracefully and ensures that the step
size does not become so large that the integrator steps over
sharp features without noticing them.

When the integrator is close to one of the boundary
conditions, it will often over-shoot. If $l$ over-shoots, we
adjust the step size so that $\Delta l = L - l$. If $\tau$
over-shoots, we adjust the step size so that $\Delta l =
(\tau - \tau_i) / \chi$ where $\chi$ is the extinction at
the mid-point of the interval. For the initial step we set
$\Delta l = \min(\tau, \tau_{\rm ch})/\chi$ where $\chi$ is
extinction at the initial point and $\tau_{\rm ch}$ is a
characteristic optical depth in the problem (often 1); if
$\chi$ is zero, we chose $\Delta l = L - l$.

We also use this integrator to determine $\tau_\infty$. We
do this simply by replacing the boundary condition on $\tau$
with one that $l$ must reach a bounding sphere enclosing the
volume in which $\chi \ne 0$. If we are not forcing
interactions, we combine steps 4 and 5 of the algorithm into
one integration by integrating until either $\tau = \tau_i$
or $l$ reaches the bounding sphere.

\subsection{Parallelism and Estimation of Uncertainty}

Monte Carlo algorithms are embarrassingly parallel and
require very little inter-process communication. Once they
have started, individual processes only need to communicate
at the end of the run to average their estimations. This
property makes Monte Carlo codes ideal for loosely-coupled
clusters of general-purpose computers connected by ordinary
networks. We have implemented parallelism using the MPICH
implementation (Gropp et al.\ 1996) of the Message Passing
Interface (MPI).

The only significant problem is generating different
sequences of pseudo-random numbers in each process. We use
distinct instances of the parallel linear-congruential
generators described by L'Ecuyer \& Andres (1996). Each
instance generates distinct sequences with periods of
$2^{71}$.

As described above, partitioning of the set of interaction
events provides a simple means to empirically gauge the
errors in the combined estimations; we have used the natural
partition that occurs when running in parallel to implement
this. Even when using only a single-processor or
dual-processor computer we often employ of order 10 parallel
processes and thereby obtain an estimate of the error in the
final results; the overhead is minimal.

\subsection{The Henyey-Greenstein Phase Function}

The phase function most commonly used for dust scattering
problems (when polarization is ignored) is the
Henyey-Greenstein phase function (Henyey \& Greenstein
1941), which is given by
\[
\Phi_\mathrm{HG}(\nu; \n, \n') 
=
(4\pi)^{-1} (1 - g_\nu^2) (1 + g_\nu^2 - 2 g_\nu \mu_s)^{-3/2},
\]
where $\mu_\mathrm{s} = \n \cdot \n'$ is the cosine
of the scattering angle $\theta_\mathrm{s}$ and the asymmetry parameter
$g_\nu$ is the mean value of $\mu_\mathrm{s}$. With this choice of
scattering function, step 8 of the algorithm is considerably
simplified, as $\mu_\mathrm{s}$ and hence the scattering angle $\theta_\mathrm{s}$
can be obtained from
\[
\mu_\mathrm{s} = \left\{
\begin{array}{ll}
1 - 2 a
&\quad \mbox{for}~g_\nu = 0\\
(2 g_\nu)^{-1}
(1 + g_\nu^2 - (1 - g_\nu^2)^2 (1 + g_\nu (1 - 2a))^{-2})
&\quad \mbox{for}~g_\nu \ne 0\\
\end{array}
\right.
,
\]
where $a$ is a deviate drawn from a uniform distribution
between 0 and 1. The azimuthal scattering angle
$\phi_\mathrm{s}$ is uniformly distributed between 0 and
$2\pi$.


\section{Test Cases}
\label{section:test-cases}

Scattering algorithms seem to be tricky to implement
correctly, and simple but non-trivial test cases are
difficult to find in the literature. For this reason, we
present two simple test cases. These test cases are
interesting in that they are non-trivial, yet the first few
intensities can be obtained by direct numerical integration.
The first test case tests the normalization of a point
source and the ratio of the unscattered and singly-scattered
intensities. The second test case tests the normalization of
a pencil beam of light and the ratios of the unscattered,
singly-scattered, and doubly-scattered intensities.

\subsection{Point Source and Slab}
\label{section:point-source-and-slab}

Consider a mono-chromatic, isotropic point source
illuminating a plane-parallel slab that is infinite in the
$x$ and $y$ directions but has a finite optical depth $T$ in
the $z$ direction. Scattering is coherent. What is
$\mathcal{L}_\infty$, the radiative intensity seen by an
observer at infinity, as a function of $\theta = \arccos
\mu$, the angle from the $+z$ direction? (We discuss the
radiative intensity in \S\ref{section:L}.)

To fix the geometry, we can take the true emissivity and
extinction distribution to be
\[
\eta_0(\r) &=& \delta(\vec r) / 4\pi,\\
\chi(z)        &=&
\left\{
\begin{array}{ll}
  \makebox[1in][l]{$T$} & \mbox{if $0 < z < 1$}\\
  \makebox[1in][l]{$0$} & \mbox{otherwise} \\
\end{array} \right. 
.
\]
This problem is similar to that considered by Henney (1998)
in his \S2.1.2, except that here we consider a point source
rather than an infinite sheet source (and we also normalize
$\Phi$ differently). Therefore, the radiative intensities
for this test case will simply be his specific intensities
multiplied by a factor of $|\mu|$ and normalized; this is a
general technique for transforming between finite unresolved
and infinite resolved sources. Thus,
$\mathcal{L}_{\infty,0}$ and $\mathcal{L}_{\infty,1}$, the
unscattered and singly-scattered emergent radiative
intensities seen from infinity, are given analytically by
\[
\mathcal{L}_{\infty,0}(\mu) &=& 
{A(\mu) \over 4\pi},\\
\mathcal{L}_{\infty,1}(\mu) &=& 
{a \mu A(\mu) \over 4\pi}
\int_{0}^{1}\!\!\!\!d\mu'
\int_0^{2\pi}\!\!\!\!d\phi\:\:
\Phi(\mu_{\rm s}(\mu,\mu',\phi)) 
{1 - e^{-T(\mu - \mu')/\mu\mu'} \over \mu - \mu'}
\]
where
\[
A(\mu) &=& 
\left\{
\begin{array}{ll}
  \makebox[1in][l]{$e^{-T/\mu}$} & \mbox{if $\mu \ge 0$}\\
  \makebox[1in][l]{$1$}          & \mbox{if $\mu < 0$} \\
  \end{array} \right. ,\\
\mu_\mathrm{s}(\mu,\mu'\!,\phi) &=& \mu\mu' +(1-\mu^2)^{1/2}(1-\mu'^2)^{1/2}
  \cos\phi .
\]

Spatially unresolved quantities seen from infinity are
independent of the precise form of the extinction
coefficient distribution within the slab, provided we
conserve the plane-parallel symmetry and the optical depth
through the slab. Therefore, when solving this problem
analytically, we can treat the slab as uniform. When solving
it with the Monte Carlo algorithm, we can choose an
extinction distribution that provides a more severe test of
the integrator, such as
\[
\chi(z) = \left\{
\begin{array}{ll} 
T\left[1/2 + \sin^2 (2\pi z)\right] &\mbox{if $0 < z < 1$}\\
0                       &\mbox{otherwise}
\end{array}
\right..
\]

\begin{figure}
\center
\includegraphics[angle=-90,width=\columnwidth]{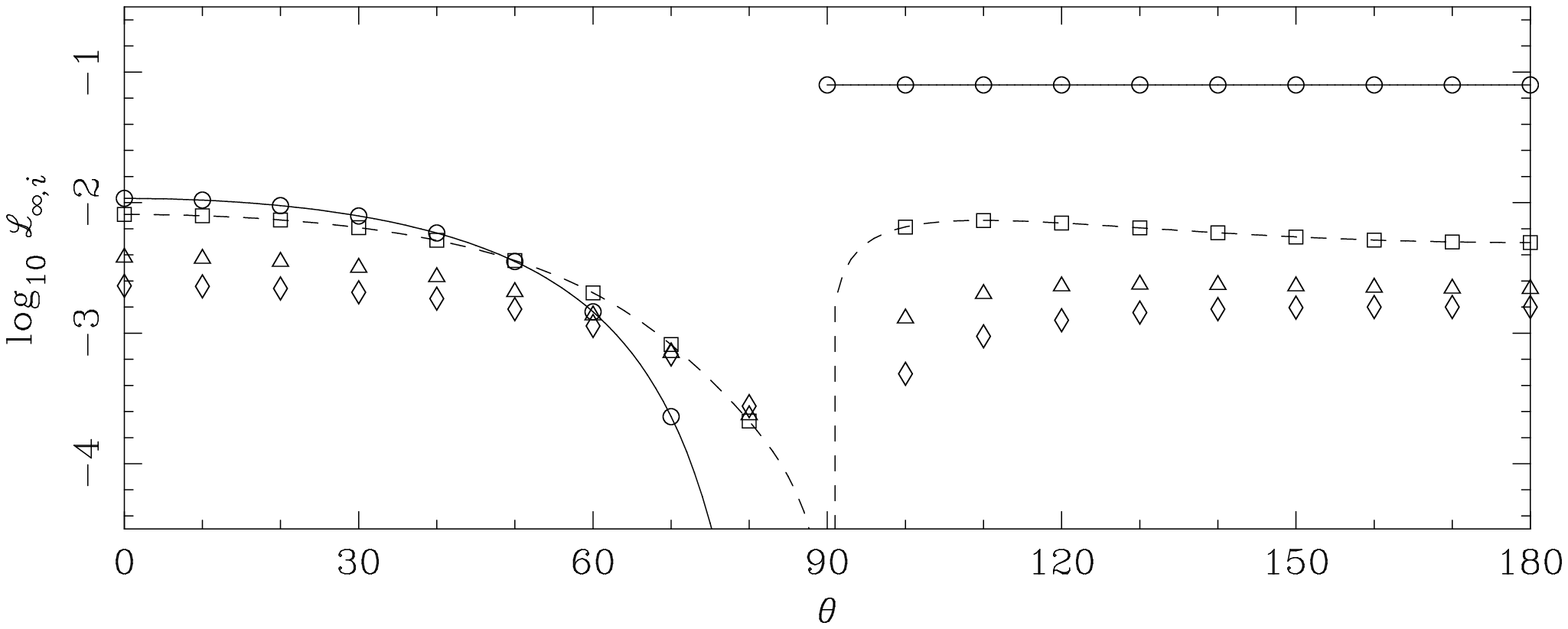}
\caption{Results for the point source and slab test case
described in \S\ref{section:point-source-and-slab}. The
figure shows the emergent radiative intensity
$\mathcal{L}_\infty(\theta)$. The lines are calculated
directly, with the solid line showing unscattered light
$\mathcal{L}_{\infty,0}$ and the dashed line showing
singly-scattered light $\mathcal{L}_{\infty,2}$. The symbols
are calculated using the Monte Carlo algorithm, with circles
$\circ$ showing the radiative intensity of unscattered light
$\mathcal{L}_{\infty,0}$, squares $\Box$ showing the
radiative intensity of singly-scattered light
$\mathcal{L}_{\infty,1}$, triangles $\triangle$ showing the
radiative intensity of double-scattered light
$\mathcal{L}_{\infty,2}$, and diamonds $\diamond$ showing
the radiative intensity of more-than-doubly-scattered light
$\mathcal{L}_{\infty,>2} \equiv \sum_{i=3}^\infty
\mathcal{L}_{\infty,i}$.
\label{figure:point-source-and-slab}}
\end{figure}

Figure~\ref{figure:point-source-and-slab} compares the
direct and Monte Carlo estimates of the emergent radiative
intensities for a slab with $T=2$, $a = 0.5$, and a
Henyey-Greenstein phase function with $g=0.5$. The values of
$\mathcal{L}_{\infty,0}$ and $\mathcal{L}_{\infty,1}$
calculated by the two methods agree to $10^{-4}$ or better,
which is commensurate with our estimates of the precisions
of the two calculations. Selected values of
$\mathcal{L}_{\infty,0}$, $\mathcal{L}_{\infty,1}$,
$\mathcal{L}_{\infty,2}$, and $\mathcal{L}_{\infty,>2}$ are
given in Table~\ref{table:point-source-and-slab}.

\subsection{Pencil Beam and Slab}
\label{section:pencil-beam-and-slab}

For our second test case, we use the same slab but replace
the point source with a pencil beam along the $+z$ axis. We
can take the true emissivity to be
\[
\eta_0(\r, \mu) = \delta(\r) \delta(\mu).
\]
Equations \ref{equation:transfer} and
\ref{equation:scattering} can be used to show that
$\mathcal{L}_{\infty,0}$, $\mathcal{L}_{\infty,1}$, and $\mathcal{L}_{\infty,2}$, the
unscattered, singly-scattered, and doubly-scattered emergent
radiative intensities seen from infinity, are given analytically by
\[
\mathcal{L}_{\infty,0}(\mu) &=& e^{-T} B(\mu)\\
\mathcal{L}_{\infty,1}(\mu) &=& 
a C(\mu)\Phi(\mu) {|\mu| \over 1 - \mu}
\left(1 - e^{-T(1 - \mu)/|\mu|}\right)
,
\label{equation:slab-singly-scattered}\\
\mathcal{L}_{\infty,2}(\mu) &=& 
a^2 
\int_0^T \!\!\!\!d\tau
\int_{-1}^{+1}\!\!\!\!d\mu'
\int_0^{2\pi}\!\!\!\!d\phi\:\:
\Phi(\mu_\mathrm{s}(\mu,\mu'\!,\phi))
\:
\Phi(\mu')
\:
e^{-\tau}
e^{-\tau'(\tau,-\mu)/|\mu|}
\:
{1 - e^{-\tau'\!(\tau, \mu')(1-\mu')/|\mu'|}
\over 1 - \mu'},
\label{equation:slab-double-scattered}
\]
where
\[
B(\mu) &=& 
\left\{
\begin{array}{ll}
  \makebox[1in][l]{$1$}&
  \mbox{if $\mu = 1$}\\
  \makebox[1in][l]{$0$}&
  \mbox{if $\mu \ne 1$}
\end{array} 
\right.
,\\C(\mu) &=& 
\left\{
\begin{array}{ll}
  \makebox[1in][l]{$e^{-T}$}&
  \mbox{if $\mu \ge 0$}\\
  \makebox[1in][l]{$1$}&
  \mbox{if $\mu < 0$}
\end{array} 
\right.
,\\
\tau'(\tau, \mu) &=& 
\left\{ 
 \begin{array}{ll}
  \makebox[1in][l]{$\tau$}      & \mbox{if $\mu \ge 0$}\\
  \makebox[1in][l]{$T - \tau$} & \mbox{if $\mu < 0$}
 \end{array} 
\right. .
\]
These results are again most easily derived using the
transformation mentioned in the previous section. Note that
as the singly-scattered light can be written in a closed
form, the doubly-scattered light is given by a single
3-dimensional integral.

\begin{table}
\caption{Results for the Point Source and Slab Test Case}
\label{table:point-source-and-slab}
\begin{center}
\begin{tabular}{ccccc}
\hline
\hline
$\theta$ &$\mathcal{L}_0(\theta)$ &$\mathcal{L}_1(\theta)$ &$\mathcal{L}_2(\theta)$ &$\mathcal{L}_{>2}(\theta)$\\
\hline
\phantom{00}0 &$1.08 \times 10^{-2}$ &$8.12 \times 10^{-3}$ &$3.80 \times 10^{-3}$ &$2.31 \times 10^{-3}$ \\
\phantom{0}10 &$1.04 \times 10^{-2}$ &$7.93 \times 10^{-3}$ &$3.73 \times 10^{-3}$ &$2.29 \times 10^{-3}$ \\
\phantom{0}20 &$9.47 \times 10^{-3}$ &$7.37 \times 10^{-3}$ &$3.53 \times 10^{-3}$ &$2.20 \times 10^{-3}$ \\
\phantom{0}30 &$7.90 \times 10^{-3}$ &$6.43 \times 10^{-3}$ &$3.19 \times 10^{-3}$ &$2.06 \times 10^{-3}$ \\
\phantom{0}40 &$5.85 \times 10^{-3}$ &$5.14 \times 10^{-3}$ &$2.70 \times 10^{-3}$ &$1.84 \times 10^{-3}$ \\
\phantom{0}50 &$3.54 \times 10^{-3}$ &$3.60 \times 10^{-3}$ &$2.07 \times 10^{-3}$ &$1.53 \times 10^{-3}$ \\
\phantom{0}60 &$1.46 \times 10^{-3}$ &$2.04 \times 10^{-3}$ &$1.37 \times 10^{-3}$ &$1.14 \times 10^{-3}$ \\
\phantom{0}70 &$2.30 \times 10^{-4}$ &$8.17 \times 10^{-4}$ &$7.07 \times 10^{-4}$ &$6.88 \times 10^{-4}$ \\
\phantom{0}80 &$7.92 \times 10^{-7}$ &$2.12 \times 10^{-4}$ &$2.38 \times 10^{-4}$ &$2.77 \times 10^{-4}$ \\
\phantom{0}90 &$7.96 \times 10^{-2}$ &0                     &0                     &0                     \\
\phantom{}100 &$7.96 \times 10^{-2}$ &$6.49 \times 10^{-3}$ &$1.30 \times 10^{-3}$ &$4.89 \times 10^{-4}$ \\
\phantom{}110 &$7.96 \times 10^{-2}$ &$7.30 \times 10^{-3}$ &$2.00 \times 10^{-3}$ &$9.47 \times 10^{-4}$ \\
\phantom{}120 &$7.96 \times 10^{-2}$ &$6.97 \times 10^{-3}$ &$2.29 \times 10^{-3}$ &$1.26 \times 10^{-3}$ \\
\phantom{}130 &$7.96 \times 10^{-2}$ &$6.40 \times 10^{-3}$ &$2.36 \times 10^{-3}$ &$1.44 \times 10^{-3}$ \\
\phantom{}140 &$7.96 \times 10^{-2}$ &$5.87 \times 10^{-3}$ &$2.34 \times 10^{-3}$ &$1.53 \times 10^{-3}$ \\
\phantom{}150 &$7.96 \times 10^{-2}$ &$5.46 \times 10^{-3}$ &$2.29 \times 10^{-3}$ &$1.57 \times 10^{-3}$ \\
\phantom{}160 &$7.96 \times 10^{-2}$ &$5.17 \times 10^{-3}$ &$2.23 \times 10^{-3}$ &$1.59 \times 10^{-3}$ \\
\phantom{}170 &$7.96 \times 10^{-2}$ &$5.00 \times 10^{-3}$ &$2.20 \times 10^{-3}$ &$1.59 \times 10^{-3}$ \\
\phantom{}180 &$7.96 \times 10^{-2}$ &$4.94 \times 10^{-3}$ &$2.18 \times 10^{-3}$ &$1.59 \times 10^{-3}$ \\
\hline
\hline
\end{tabular}
\end{center}
\end{table}

\begin{figure}[p]
\center
\includegraphics[angle=-90,width=\columnwidth]{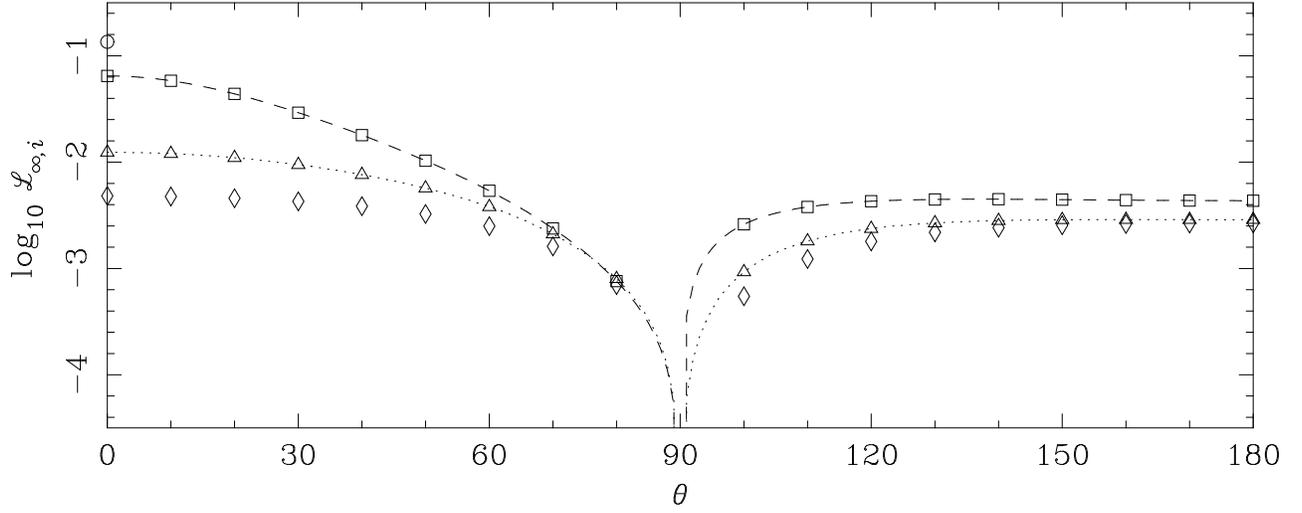}
\caption{Results for the pencil beam and slab case described
in \S\ref{section:pencil-beam-and-slab}. The figure shows
the emergent radiative intensity $\mathcal{L}_\infty(\theta)$. The lines are
calculated directly, with the dashed line showing
singly-scattered light $\mathcal{L}_{\infty,1}$ and the dotted line
showing doubly-scattered light $\mathcal{L}_{\infty,2}$. The symbols
are calculated using the Monte Carlo algorithm, with circles
$\circ$ showing the radiative intensity of unscattered light
$\mathcal{L}_{\infty,1}$, squares $\Box$ showing the radiative intensity of
singly-scattered light $\mathcal{L}_{\infty,2}$, triangles $\triangle$
showing the radiative intensity of doubly-scattered light
$\mathcal{L}_{\infty,2}$, and diamonds $\diamond$ showing the
radiative intensity of more-than-doubly-scattered light $\mathcal{L}_{\infty,>2}
\equiv \sum_{i=3}^\infty \mathcal{L}_{\infty,i}$.
\label{figure:pencil-beam-and-slab}}
\end{figure}

\begin{table}
\caption{Results for the Pencil Beam and Slab Test Case}
\label{table:pencil-beam-and-slab}
\begin{center}
\begin{tabular}{ccccc}
\hline
\hline
$\theta$ &$\mathcal{L}_0(\theta)$ &$\mathcal{L}_1(\theta)$ &$\mathcal{L}_2(\theta)$ &$\mathcal{L}_{>2}(\theta)$\\
\hline
\phantom{00}0 &$1.35 \times 10^{-1}$ &$6.46 \times 10^{-2}$ &$1.24 \times 10^{-2}$ &$4.79 \times 10^{-3}$ \\
\phantom{0}10 &0                     &$5.82 \times 10^{-2}$ &$1.20 \times 10^{-2}$ &$4.74 \times 10^{-3}$ \\
\phantom{0}20 &0                     &$4.39 \times 10^{-2}$ &$1.10 \times 10^{-2}$ &$4.57 \times 10^{-3}$ \\
\phantom{0}30 &0                     &$2.92 \times 10^{-2}$ &$9.45 \times 10^{-3}$ &$4.27 \times 10^{-3}$ \\
\phantom{0}40 &0                     &$1.80 \times 10^{-2}$ &$7.63 \times 10^{-3}$ &$3.84 \times 10^{-3}$ \\
\phantom{0}50 &0                     &$1.03 \times 10^{-2}$ &$5.69 \times 10^{-3}$ &$3.26 \times 10^{-3}$ \\
\phantom{0}60 &0                     &$5.38 \times 10^{-3}$ &$3.80 \times 10^{-3}$ &$2.51 \times 10^{-3}$ \\
\phantom{0}70 &0                     &$2.37 \times 10^{-3}$ &$2.11 \times 10^{-3}$ &$1.61 \times 10^{-3}$ \\
\phantom{0}80 &0                     &$7.60 \times 10^{-4}$ &$7.98 \times 10^{-4}$ &$7.02 \times 10^{-4}$ \\
\phantom{0}90 &0                     &0                     &0                     &0                     \\
\phantom{}100 &0                     &$2.60 \times 10^{-3}$ &$9.25 \times 10^{-4}$ &$5.44 \times 10^{-4}$ \\
\phantom{}110 &0                     &$3.78 \times 10^{-3}$ &$1.81 \times 10^{-3}$ &$1.22 \times 10^{-3}$ \\
\phantom{}120 &0                     &$4.29 \times 10^{-3}$ &$2.37 \times 10^{-3}$ &$1.79 \times 10^{-3}$ \\
\phantom{}130 &0                     &$4.46 \times 10^{-3}$ &$2.67 \times 10^{-3}$ &$2.18 \times 10^{-3}$ \\
\phantom{}140 &0                     &$4.48 \times 10^{-3}$ &$2.82 \times 10^{-3}$ &$2.41 \times 10^{-3}$ \\
\phantom{}150 &0                     &$4.44 \times 10^{-3}$ &$2.87 \times 10^{-3}$ &$2.55 \times 10^{-3}$ \\
\phantom{}160 &0                     &$4.39 \times 10^{-3}$ &$2.88 \times 10^{-3}$ &$2.61 \times 10^{-3}$ \\
\phantom{}170 &0                     &$4.35 \times 10^{-3}$ &$2.88 \times 10^{-3}$ &$2.64 \times 10^{-3}$ \\
\phantom{}180 &0                     &$4.34 \times 10^{-3}$ &$2.88 \times 10^{-3}$ &$2.65 \times 10^{-3}$ \\
\hline
\hline
\end{tabular}
\end{center}
\end{table}

Figure~\ref{figure:pencil-beam-and-slab} compares the direct
and Monte Carlo estimates of the emergent radiative
intensities for a slab with $T=2$, $a = 0.5$, and a
Henyey-Greenstein phase function with $g=0.5$. The values of
$\mathcal{L}_{\infty,1}$ and $\mathcal{L}_{\infty,2}$
calculated by the two methods agree to $5 \times 10^{-4}$ or
better, which is commensurate with our estimates of the
precisions of the two calculations. Selected values of
$\mathcal{L}_{\infty,0}$, $\mathcal{L}_{\infty,1}$,
$\mathcal{L}_{\infty,2}$, and $\mathcal{L}_{\infty,>2}$ are
given in Table~\ref{table:pencil-beam-and-slab}.


\section{The Importance of Forced Scatterings and Interactions}
\label{section:optimizations}

\begin{figure}
\center
\vspace{6mm}
\includegraphics[angle=-90,width=\columnwidth]{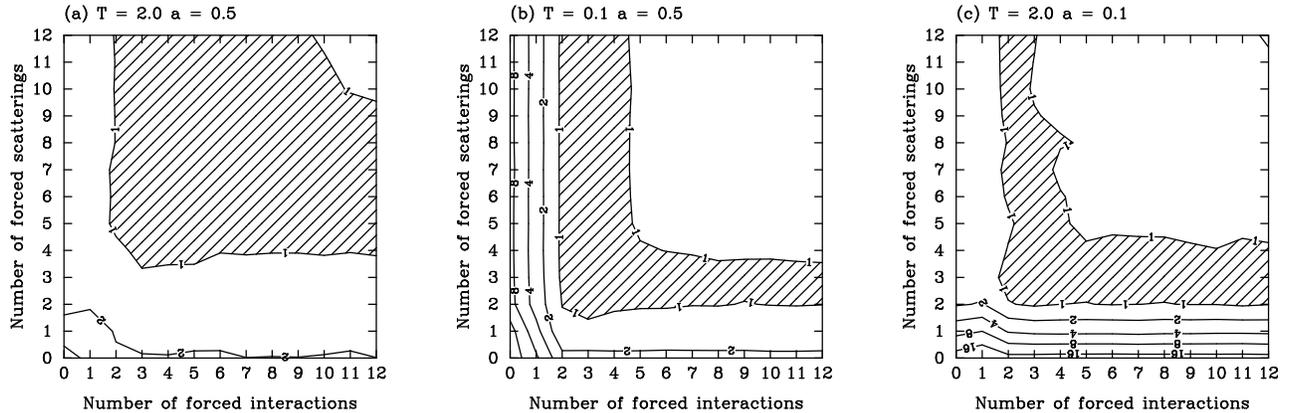}
\vspace{6mm}
\caption{Contour plots of the relative errors in three
trials of the slab geometry as functions of the number of
forced scatterings and forced interactions. Trial (a) has $T
= 2$ and $a = 0.5$, trial (b) has $T = 0.1$ and $a = 0.5$,
and trial (c) has $T = 2.0$ and $a = 0.1$; in all cases $g =
0.5$ and $\theta = 45^\circ$. Contours are spaced by factors
of 2. The hatched regions have the lowest errors.
\label{figure:error}}
\end{figure}

To illustrate the importance of forced scatterings
(\S\ref{section:forced-scatterings}) and forced interactions
(\S\ref{section:forced-interactions}), we have examined the
relative errors in the emergent intensity in simulations of
the slab geometry (\S\ref{section:test-cases}) as functions
of the number of forced scatterings and the number of forced
interactions. In order to make the comparison fair, we ran
each calculation for the same amount of processor time. The
relative errors were estimated by partitioning the sample of
pseudo-photons into 48 sub-samples.

We ran three trials with different values of $T$ and $a$.
Trial (a) has $T = 2.0$ and $a = 0.5$, the same as the test
case considered in \S\ref{section:test-cases}, and is
moderately optically-thick, with singly-scattered and
doubly-scattered light dominating more-than-doubly-scattered
light. Trial (b) has $T = 0.1$ and $a = 0.5$, and is
optically-thin, with singly-scattered light dominating
multiply-scattered light. Trial (c) has $T = 2.0$ and $a =
0.1$, and is moderately optically-thick, but has a low
albedo, so singly-scattered light also dominates
multiply-scattered light. In all cases we kept $\theta =
45^\circ$ and $g=0.5$.

Figure \ref{figure:error} shows contour plots of the
relative errors for these trials. The best results for trial
(a), with $T=2$ and $a=0.5$, are obtained in a broad region
with at least 4 forced scatters and 3 forced interactions.
The best results for trial (b), with $T=0.1$ and $a=0.5$,
were obtained for at least 2 forced interactions and at
least 2 forced scatters. The best results for trial (c),
with $T=2.0$ and $a=0.1$, were obtained for at least 2
forced interactions and at least 3 forced scatters. In all
cases, however, the optimal region is roughly L-shaped. That
is, excessive numbers of either forced scatters or forced
interactions have little effect, but excessive numbers of
both start to increase the error again.

The best errors for trials (a), (b), and (c) are roughly 7,
90, and 30 times smaller than for a naive implementation
with no forced scatters or forced interactions Since the
relative error in Monte Carlo calculations decreases as the
square root of the computational effort, these optimizations
represent savings in time of factors of roughly 50, 8000,
and 1000.

These results can be understood qualitatively as follows. On
average, the effect of forcing scatterings and interactions
is to keep the pseudo-photon in the system for longer, which
has two effects: (i) the pseudo-photon is more likely to
contribute to derived quantities that depend on scattered
light, which reduces the variance in these quantities, and
(ii) the computational cost per pseudo-photons increases, so
fewer can be followed, which increases the variance in
derived quantities. In the cases examined here, the former
is more important to begin with; for moderate numbers of
forced scatterings and interactions, making each
pseudo-photon more useful compensates for the smaller total
number of pseudo-photons, and the variance is reduced. This
is especially important if the system is optically thin
(trial b) or if the albedo is low (trial c), as
pseudo-photons can easily leave the system or be absorbed,
and thereby make relatively little contribution to derived
quantities.

However, if the number of both forced scatterings and
interactions is excessive, each pseudo-photon is forced to
remain in the system for many scatterings. This may reduced
the variance of quantities derived from very highly
scattered light, but it reduces the total number of
pseudo-photons that can be considered, and thereby increases
the variance of derived quantities that are dominated by
moderately scattered light. Furthermore, the optimal region
is L-shaped because, for the optical depths and albedos
considered here, \emph{both} excessive numbers of forced
scatterings \emph{and} excessive numbers of forced
interactions are required to keep each pseudo-photons in the
system for many scatterings; if one or the other is not
excessive, the pseudo-photon is absorbed or escapes. 

It is impossible to give a universally applicable rule to
select a priori the optimal number of forced scatterings and
interactions, as these quantities depend on the geometry and
also on the particular quantity being calculated. However,
the discussion in the previous paragraph implies that the
optimal values for both will probably be roughly equal. Our
experience suggests that two to four forced scatterings and
interactions might be a useful rule of thumb; this number is
not too far from optimal for any of the cases we have
investigated.

Naive Monte Carlo algorithms are often considered ill-suited
to optically-thin problems, as the overwhelming majority of
photons escape without interacting. However, is not the case
for more sophisticated algorithms that force the first few
interactions, as our optically-thin trial demonstrates. A
specialized single-scattering plus attenuation calculation
would probably calculate the emergent singly-scattered
intensity more efficiently, but a Monte Carlo calculation
may well be more flexible and can directly account for
multiply-scattered light.


\section{Time-Dependence and Polarization}
\label{section:extensions}

The algorithm as presented covers the time-independent
transfer of unpolarized light. Both of these restrictions
can be lifted quite easily. 

The simplest way of including general time-dependence is to
sample the range of times of emission, keep track of the
travel time of the pseudo-photon within the system, and
account for the travel time when computing quantities of
interest. If only the source varies, it is probably more
efficient to construct a ``response function'' for each
quantity of interest which can be convolved with changes in
the brightness of the source to predict the time-variation
of that quantity.

Perhaps the simplest way to include polarization is to
generalize the scalar statistical weight $w$ to a vector
statistical weight $\vec w \equiv (w_I, w_Q, w_U, w_V)$,
whose components correspond to the four components of the
Stokes vector. The components will in general not be
conserved upon scattering, but will transform according to
the adopted scattering matrix. See, for example, Hillier
(1991) and Whitney (1991).


\section{Summary}

We have presented and discussed a Monte Carlo algorithm for
a restricted class of scattering problems. Our main
contributions have been to formalize this algorithm, to
present two test cases, and to investigate its efficiency.
Important features of the algorithm are forced escapes,
forced scatterings, and forced interactions. Using forced
escapes to estimate the emergent intensity can be orders of
magnitude more efficient than naively binning escaping
pseudo-photons. Also, using a small number (two to four) of
forced scatterings and forced interactions can significantly
improve the efficiency of the algorithm for many problems,
overwhelmingly so for optically-thin problems or those where
the albedo is small.


\acknowledgements

We are grateful to Karl Stapelfeldt and John Krist for
serving as crash-test dummies for several years, to Jon
Bjorkman, Kenny Wood, and Barbara Whitney for many useful
discussions on Monte Carlo techniques and applications, and
to an anonymous referee. This work was supported by CONACyT
project 27570E.


\end{document}